\documentclass[12pt]{article}%

\begin{document}
\begin{center}{\large\bf Fundamental Quantal Paradox and Its Resolution}\end{center}

\vskip 1em \begin{center} {\large Felix M. Lev} \end{center}

\vskip 1em \begin{center} {\it Artwork Conversion Software Inc.,
1201 Morningside Drive, Manhattan Beach, CA 90266, USA
(Email:  felixlev314@gmail.com)} \end{center}

\begin{abstract} 
The postulate that coordinate and momentum representations are related
to each other by the Fourier transform has been accepted from the beginning of quantum theory.   
As a consequence, coordinate wave functions of 
photons emitted by stars have cosmic sizes. This results in a paradox because predictions
of the theory contradict observations. The reason of the paradox and its resolution are
discussed.
\end{abstract}

\begin{flushleft} PACS: 03.65.-w, 03.65.Sq, 03.65.Ta \end{flushleft}

\begin{flushleft} Keywords: quantum theory, position operator, semiclassical approximation\end{flushleft}

\section{Introduction}
\label{intro}

Although quantum theory exists for a rather long period of time, the problem of its 
interpretation is still widely debated. Let us first make a remark about the terminology of the theory. The term "wave function" has arisen at the beginning of 
quantum era in efforts to explain quantum behavior in terms of classical waves but now it is clear
that no such explanation exists. The notion of wave
is purely classical; it has a physical meaning only as a way of describing systems of many particles by
their mean characteristics. The problem discussed in the present paper is a good illustration
that the term "wave function" might be misleading
since in quantum theory it defines not amplitudes of waves but amplitudes of probabilities. 
So, although in our opinion the term "state vector" is more pertinent than "wave function" we use the
latter in accordance with the usual terminology. 

For example, the electron has an electric charge $e$ which is
indivisible, i.e. roughly speaking the electron is a point. The electron wave function (WF) $\psi({\bf r})$ 
describes only probabilities to find the electron {\it as a whole} at different points ${\bf r}$.
Therefore $e|\psi({\bf r})|^2$ cannot be treated as the charge density in the
classical sense. If a decomposition $\psi=\psi_1+
\psi_2$ is possible according to the superposition principle, this does not mean splitting the
electron into two parts with the charges $e_1$ and
$e_2$ such that $e_1+e_2=e$.  Therefore classical description of elementary particles is not
adequate. Those remarks will be important in Sec. \ref{paradox}.

Note also that standard quantum theory is based on 
classical mathematics involving infinitely small, infinitely large, continuity etc. The part of this mathematics 
which is used in quantum theory
was developed mainly when people believed that every object can be divided 
into arbitrarily large numbers of arbitrarily small parts. However, in view of the existence of
elementary particles, this is not the case. Hence a problem arises whether classical mathematics 
is pertinent on fundamental quantum level. In Refs. \cite{GFQT,DS} and references 
therein we argue that fundamental quantum theory should be based on finite mathematics. However,
in the present paper we proceed from standard quantum theory.

One might say that the problem of coordinate description of elementary particles does not
arise because {\it the results} of existing fundamental quantum theories 
describing interactions on quantum level (QED, electroweak theory and QCD) are formulated 
exclusively in terms of the S-matrix in momentum space without mentioning space-time at all.

For example, consider the problem of calculating the hydrogen energy levels. 
As follows from Feynman diagrams for the one-photon exchange, in the approximation $(v/c)^2$  the electron in the hydrogen atom 
can be described in the potential formalism where the potential acts on the WF in momentum space.  So for calculating energy 
levels one should solve the eigenvalue problem for the Hamiltonian with this potential. This is an integral equation which can 
be solved by different methods. One of them is to apply the Fourier transform and get standard Schr\"{o}dinger or Dirac 
equation in coordinate representation. Hence the fact that the results for energy levels are in good agreement with experiment 
shows that QED defines the potential correctly and {\it standard coordinate Schr\"{o}dinger and Dirac equations are only convenient mathematical ways of solving the eigenvalue problem}. For this problem the physical meaning of the position operator is not important. One can consider other transformations of the original integral equation and define other position operators. 

However,  since quantum theory is treated as a more general than classical one, in semiclassical approximation it should reproduce the motion of a particle along
the classical trajectory defined by classical equations of motion. Hence the position operator 
is needed only in semiclassical approximation and it should be {\it defined} from additional considerations. 

In standard approaches to quantum theory the existence of space-time background is assumed from the beginning. Then the position operator for a particle in this background is the operator of multiplication by the particle radius-vector ${\bf r}$. A problem arises how to define the particle momentum
operator. In the literature the choice $-i\hbar \partial/\partial{\bf r}$ for this operator is justified from
several considerations. Below we briefly describe standard reasons for such a choice. A more
detailed discussion can be found in subsection 1.1 of Ref. \cite{lev1} titled "Historical reasons for choosing standard form of position operator".

From the beginning of quantum theory it has been postulated by analogy with classical 
electrodynamics that the coordinate and momentum representations are related to each other by the Fourier transform.  However, the interpretations
of classical fields on one hand and wave functions on the other are fully different
and such a situation does not seem to be natural.

Another reason is that, as explained in textbooks (see e.g. Ref. \cite{LLIII}), the above result 
for the momentum operator can be justified from the requirement that quantum theory should correctly reproduce classical results in semiclassical approximation. 
However, this requirement does not define the operator unambiguously. Indeed, if the operator
$B$ becomes zero in semiclassical limit then the operators $A$ and $A+B$ have the same
semiclassical limit. 

Yet another reason follows. As assumed by Dirac \cite{Dirac}, the commutator of the momentum and
position operators is proportional to
the corresponding classical Poisson bracket with the coefficient $i\hbar$. Then, as follows
from the Stone-von Neumann theorem, the above result for the momentum operator is
valid up to unitary equivalence. However, this
argument is not convincing because there are no physical reasons to require that the commutator 
is a c-number. 

 As argued in Ref. \cite{DS}, quantum theory should start not from
classical space-time background but from a symmetry algebra. For example, on quantum level Poincare invariance means
not that the space-time background is Minkowskian but that operators describing a system satisfy the
commutation relations of the Poincare algebra. Then elementary particles are described by irreducible
representations (IRs) of this algebra. Such IRs can be implemented in spaces of functions 
$\chi({\bf p})$ such that
\begin{equation}
||\chi||^2=\int |\chi({\bf p})|^2d^3{\bf p}<\infty
\label{normchi}
\end{equation}
where $||...||$ is the norm and the momentum operator is the operator of multiplication by ${\bf p}$. At the same time, among
the ten independent operators in such IRs there is no position operator.

One of the conditions characterizing semiclassical approximation is that the orbital angular momentum of
the particle under consideration should be large. Then, as follows from explicit expressions for
the representation operators (see e.g. Ref. \cite{JPA}), in this approximation spin effects can be
neglected. A standard assumption is that the position operator in the representation (\ref{normchi}) is
$i\hbar\partial/\partial {\bf p}$. One of the arguments is that if the coordinate
wave function contains $exp[iS({\bf r},t)/\hbar]$ where $S({\bf r},t)$ is the classical action then the 
Schr\"{o}dinger equation becomes
the Hamilton-Jacobi one in the formal limit $\hbar\to 0$.

With this choice, an inevitable consequence of quantum theory is the effect of wave packet
spreading (WPS) discussed in textbooks and many papers. 
At the beginning of quantum theory the WPS effect has been investigated by de Broglie,
Darwin and Schr\"{o}dinger. The fact that WPS is inevitable has been treated by several authors as unacceptable. For example, de Broglie has proposed to
describe a free particle not by the Schr\"{o}dinger equation but by a wavelet which satisfies a nonlinear equation
and does not spread out (see e.g. Ref. \cite{Barut}).

At the same time, it has not been explicitly shown that WPS is 
incompatible with experimental data. For example, as shown by Darwin \cite{Darwin}, for macroscopic
bodies this effect is negligible. Probably it is also believed that in 
experiments on the Earth with atoms 
and elementary particles spreading does not have enough time to manifest itself. 
Probably for those reasons the absolute majority of physicists do not treat WPS as a drawback of the theory. 

However, a natural question arises what happens 
to photons which can travel from distant objects to Earth even for billions of years. 
To answer this question one needs to know the photon position operator. This problem has been
discussed by many authors. Some of them state that the photon coordinate wave function 
does not exist. For example, the argument in Ref. \cite{AB} is that if the photon coordinate WF is defined in standard way then the photon electric and magnetic fields at any point will depend
not on the WF at this point but on integrals from the WF over a region with the
dimension of the photon wave length. 

This argument is problematic for several reasons. For example, as noted by Pauli (see p. 191 of Ref.
\cite{Pauli}), the photon coordinates cannot have a physical meaning at distances less than the wave length.
Since the wave length is defined as $\lambda=2\pi\hbar/p$ where $p$ is the magnitude
of momentum, then $\lambda\to 0$ in the formal limit $\hbar\to 0$. Therefore on semiclassical level
any WFs differing only by the behavior at distances less than the wave length are equivalent.

The fact that the photon coordinate wave function should exist follows from simple considerations.
For example, if a photon emitted on Sirius flies to Earth then the theory should say, at least approximately, where this photon at different moments of time is: in the vicinity of Sirius,
on its half-way, near the Earth etc. In numerous papers on the photon coordinate wave function
the existence or nonexistence of this function is discussed in conjunction with the electromagnetic
field of a single photon. However, in quantum theory the description of a free particle is fully defined by its IR depending 
only on the mass and spin. It does not depend on interactions this particle
participates in, on whether the particle is neutral or has antiparticles etc.. As noted above, 
the coordinate description of elementary particles
is needed only in semiclassical approximation and in this 
 approximation the particle spin is not important. 
{\it Hence the description of a free ultrarelativistic particle does not depend on
the nature of the particle, i.e. on whether the particle is the photon, the proton, the electron etc.} 

Relativistic position operator has been discussed by many authors and in all those 
publications it has been assumed that the the coordinate and momentum spaces are
related to each other by the Fourier transform. The first consideration of this operator
has been given by Newton and Wigner \cite{NW}. Their result
is that if spin effects are neglected then the form of the position operator in the representation 
(\ref{normchi}) is the same as in nonrelativistic case, i.e. $i\hbar\partial/\partial {\bf p}$. 
Position operators discussed by other authors differ from the Newton-Wigner one only by spin terms and
by the behavior at distances less than the wave length (see Refs. \cite{Keller,lev1} for a more detailed
discussion). Therefore in semiclassical approximation all those operators are equivalent. 

Some authors require that in the relativistic case the coordinate description should be given
only in the covariant form, i.e. by using only four-vectors. Strictly speaking this is not necessary
since relativistic invariance means only that the operators commute according to the
commutation rules in the Poincare algebra. Nevertheless, the covariant coordinate description
can be implemented as follows.

We first note that the relativistic coordinate wave function $\Psi({\bf r}, t)$ cannot be defined in the whole Minkowski space. Usually it is defined at some fixed value of $t$. In general it can be
defined on a space-like hypersurface of Minkowski space defined by the condition $(n,x)=\tau$
where $n$ is a time-like four-vector and $x$ is an arbitrary vector in Minkowski space.
Then, as discussed in Ref. \cite{lev1}, the coordinate description can be given in the covariant
form. In the present paper we discuss only coordinate wave functions defined on hypersurfaces
$t=const$ when the four-vector $n$ has the component $n=(1,0,0,0)$. Then the spinless Newton-Wigner position operator is $i\hbar\partial/\partial {\bf p}$. 

For all those reasons, in Sec. \ref{WF} we calculate the coordinate WF of an ultrarelativistic particle
proceeding from the usual assumption that the coordinate and momentum representations are related to
each other by the Fourier transform. As argued in Sec. \ref{paradox}, the results contradict the usual experience
on how we observe stars. The reason of the paradox and its resolution are discussed in Sec. \ref{reason}. 

\section{Coordinate wave function of ultrarelativistic particle}
\label{WF}

Consider a model where the momentum WF is
\begin{equation}
\chi({\bf p})=f({\bf p}/p)F(p)/p
\label{relchipn}
\end{equation}
Let $f({\bf p}/p)=\sum_{l\mu}c_{l\mu}Y_{l\mu}({\bf p}/p)$ be the
decomposition of the function $f$ over spherical functions.  Since the energy of the ultrarelativistic particle is $E=pc$, the dependence of this function on $t$ is given by
\begin{equation}
\chi({\bf p},t)=exp(-\frac{i}{\hbar}pct)\chi({\bf p})
\label{chit}
\end{equation}
Since we assume that the momentum and coordinate WFs are related to each
other by the Fourier transform, the coordinate WF is
\begin{equation}
\psi({\bf r},t)=\int exp(\frac{i}{\hbar}{\bf p}{\bf r})\chi({\bf p},t)\frac{d^3{\bf p}}{(2\pi\hbar)^{3/2}}
\label{Fourier}
\end{equation}

For calculating this function we use the known decomposition:
\begin{equation}
exp(\frac{i}{\hbar}{\bf pr})=4\pi\sum_{l\mu}i^lj_l(pr/\hbar )Y_{l\mu}^*({\bf p}/p)Y_{l\mu}({\bf r}/r)
\label{flat}
\end{equation}
where $r=|{\bf r}|$ and $j_l$ is the spherical Bessel function. Its asymptotic expression when the argument is large is $j_l(x)\approx sin(x-\pi l/2)/x$. Then, as follows from the orthogonality 
condition for
spherical functions and Eqs. (\ref{relchipn}-\ref{flat}), at large distances
\begin{eqnarray}
&&\psi({\bf r},t)=\frac{-i}{(2\pi\hbar)^{1/2}r}\sum_{l\mu}c_{l\mu}Y_{l\mu}({\bf r}/r)
[G(ct-r)-(-1)^lG(ct+r)]
\label{relpsirn}
\end{eqnarray}
where
\begin{equation}
G(\xi)=\int_0^{\infty} F(p)exp(-\frac{i}{\hbar}\xi p )dp
\label{Gxi}
\end{equation}

For reasonable choices of $F(p)$ we will have that at large distances and times $G(ct-r)\gg G(ct+r)$. 
Indeed if, for example, the quantities $p_0$ and $a$ are such that 
$p_0>0$, $a>0$ and $p_0a\gg \hbar$ then possible $(F,G)$ choices are: 
\begin{eqnarray}
&&F(p)=exp(-\frac{|p-p_0|a}{\hbar}),\quad G(\xi)=\frac{exp(-ip_0\xi/\hbar)}{a^2+\xi^2};\nonumber\\
&&F(p)=exp(-\frac{(|p-p_0|a)^2}{2\hbar^2}),\quad 
G(\xi)=(2\pi)^{1/2}\frac{\hbar}{a}exp(-\frac{ip_0\xi}{\hbar}-\frac{\xi^2}{2a^2})
\label{FG}
\end{eqnarray}
Then, as follows from Eq. (\ref{relpsirn}), in those cases
\begin{eqnarray}
\psi({\bf r},t)=\frac{-i}{(2\pi\hbar)^{1/2}r}f({\bf r}/r)G(ct-r)
\label{relpsif}
\end{eqnarray}

Therefore at large distances and times the coordinate WF has the following
properties: a) at each moment of time $t$ the radial part is not negligible only
inside a thin sphere with the radius $ct$ and the width of the order of $a$; b) the 
angular part does not depend on time and is the same as the angular part
of the momentum wave function. As shown in Refs. \cite{Naumov,lev1} and references therein, similar conclusions can be obtained for another models of the particle WF. Therefore those conclusions are rather general.

According to the present knowledge, photons emitted by stars belong to two categories.
The first one consists of photons emitted from transitions between different atomic energy levels. For example,
the most known case is the Lyman transition $2P\to 1S$ in the hydrogen atom. For such photons the orbital
angular momentum is small, i.e. the sphere spreads almost uniformly in all directions. In this case the
WF can be qualitatively described in the spherically symmetric approximation when $f=const$. 

However, the main part of the radiation
can be approximately described in the blackbody model such that typical photon energies 
are not close to energies of absorption lines
for that matter. Then the photon WFs can be treated as wave packets (see Ref. \cite{lev1}
for a more detailed discussion). If the mean value of the momentum is directed along the positive
direction of the $z$-axis then the function $f$ is essentially different from zero only in the range where
angles between momenta and the $z$-axis are small. If $p_0$ is the magnitude of the mean
momentum and $p_{\bot}$ is the uncertainty of the transverse momentum such
that $p_{\bot}\ll p_0$ then $\alpha=p_{\bot}/p_0$ is the angular uncertainty. Since the angular
dependences of the coordinate and momentum WFs are the same, the transversal width
of the packet spreads out with time as $\alpha ct$.

The estimation of typical values of $\alpha$ is problematic because photons can be created
in different processes. However, even if $\alpha$ is small, after years or longer
the quantity $\alpha ct$ becomes much greater than sizes of stars and planets (see estimations
in Ref. \cite{lev1}). So,  as a consequence of WPS, photons emitted by stars have
WFs of cosmic sizes and the motion of such photons is not a motion along classical trajectories. 

A problem arises whether this phenomenon is compatible with our experience in observations of
stars. We will discuss this problem in Sec. \ref{paradox}. However, at least if the approximation
leading to the derivation of Eq. (\ref{relpsif}) is valid then images of stars will not appear blurred. Indeed, suppose that we observe a star which has the radius
$R$ and is on the distance $L$ from Earth. Then the image of the star will not appear blurred if
the angular resolution is or the order of $R/L$ of better. For all stars the quantity
$R/L$ is typically much less than $\alpha$. Suppose that a 
photon is created at the origin and Earth is seen from the origin in the narrow angular range
defined by the function ${\tilde f}({\bf r}/r)$. Suppose that the support of ${\tilde f}({\bf r}/r)$
is within the range defined by $f({\bf r}/r)$. Then the projection of the WF (\ref{relpsif})
onto Earth is given by the same expression where $f({\bf r}/r)$ is replaced by 
${\tilde f}({\bf r}/r)$. Since the angular WFs in coordinate and momentum 
representations are the same, the momenta  measured on Earth will be in the angular range
defined by the function ${\tilde f}({\bf p}/p)$. This result is the generalization of the solution
of the Mott-Heisenberg problem (see e.g. Ref. \cite{lev1}). At the same time, as argued in 
Sec. \ref{paradox}, the fact that photons emitted by stars have WFs with cosmic sizes does
lead to a paradox in observations of stars. 

\section{Does light from stars consist of free photons?}

The answer to this question depends on: 1) whether direct interaction between the photons is
important; 2) whether their interaction with the interstellar medium is important and 
3)whether
coherent photon states play an important role in the star radiation. As noted in the
literature (see e.g. Ref. \cite{lev1}), at energies typical in the star radiation the cross section of 
the photon-photon interaction is so small that the role of this interaction is negligible.

As far as item 2) is concerned, one can note the following.  
The problem of explaining the redshift phenomenon has a long history. Different competing approaches can be 
divided into two big sets which we call Theory A and Theory B. In Theory A the redshift has been originally explained 
as a manifestation of the Doppler effect but in recent years  the cosmological and gravitational redshifts 
have been added to the consideration. In this theory the interaction of photons with the interstellar medium is treated as
practically not important. On the contrary, in Theory B, which is often called the tired-light theory, the interaction of photons with the interstellar medium is treated as the main reason for the redshift. The majority of physicists 
believe that Theory A explains the astronomical data better than Theory B because  
any sort of scattering of light would predict more blurring than is seen (see e.g.
the article "Tired Light" in Wikipedia). 
As follows from these remarks, in Theory A it is assumed that with a good accuracy we can treat photons 
as propagating in empty space. 

We now consider item 3). In fundamental quantum theories elementary particles are described 
by states in the Fock space. 
Let $a({\bf p},\lambda)$ and $a({\bf p},\lambda)^*$ be the annihilation and creation operators for a
photon with the momentum ${\bf p}$ and polarization $\lambda$. They satisfy the 
commutation relations
\begin{eqnarray}
&&[a({\bf p}',\lambda'),a({\bf p}'',\lambda'')]=
[a({\bf p}',\lambda')^*,a({\bf p}'',\lambda'')^*]=0,\nonumber\\
&&[a({\bf p}',\lambda'),a({\bf p}'',\lambda'')^*]=
\delta^{(3)}({\bf p}''-{\bf p}')\delta(\lambda''-\lambda')
\label{commutatorA}
\end{eqnarray}
and if $\Phi_0$ is the vacuum state then $a({\bf p},\lambda)\Phi_0=0$. 
A general form of the state vector of the electromagnetic field is
\begin{eqnarray}
\Phi(t)=\sum_{n=0}^{\infty}\sum_{\lambda_1,...\lambda_n}\int...\int 
\chi_n({\bf p}_1,\lambda_1,...{\bf p}_n,\lambda_n)
\prod_{i=1}^n \{exp(-\frac{i}{\hbar}|{\bf p}_i|ct)a({\bf p}_i,\lambda_i)^*
 d^3{\bf p}_i \}\Phi_0
\end{eqnarray}
where $\chi_n({\bf p}_1,\lambda_1,...{\bf p}_n,\lambda_n)$ can be called the
WF of the $n-$photon state.

Two extreme cases of the state vector are as follows. The first case, which can be
called strongly incoherent, is such that the functions 
$\chi_n({\bf p}_1,\lambda_1,...{\bf p}_n,\lambda_n)$ have sharp maxima at
${\bf p}_i={\bf p}_i^0\,\, (i=1,...n)$ and all the values ${\bf p}_i^0$ are considerably
different. The second case, which can be called strongly coherent, is such that
\begin{equation}
\chi_n({\bf p}_1,\lambda_1,...{\bf p}_n,\lambda_n)=c_n\prod_{i=1}^n\chi({\bf p}_i,\lambda_i)
\label{coherent}
\end{equation}
where the function $\chi({\bf p},\lambda)$ is different from zero only for one value of $\lambda$
and has a sharp maximum at ${\bf p}={\bf p}^0$ while the coefficients $c_n$ are chosen from
some special consideration.  

The density of radiation coming to us from distant stars is very small. Therefore the
assumption  that this radiation can be described in the framework of classical
electrodynamics is problematic and one might think that this radiation can be treated simply as a
collection of independent photons. However, one of the arguments in favor of this assumption is the
Hanbury Brown and Twiss experiment. Here two photomultiplier tubes separated by about 6 meters, were aimed at Sirius and excellent angular resolution has been achieved. The theoretical explanation of the
experiment (see e.g. Ref. \cite{Scully}) can be given both, in terms of classical optics and
in terms of interference of independent photons (which are not in coherent states). 
In the latter case it is also assumed that radiation from Sirius has the blackbody type.

However, Sirius is the brightest star on the sky and its distance to Earth is "only" 8.6 light years.
The angular resolution of Sirius is at the limit of modern telescope arrays and this resolution is
insufficient for determining radii of other stars.  Conclusions about them are made from the data on luminosity and temperature assuming that the major part of the radiation can be described in
the blackbody model. Therefore even if a conclusion about radiation from Sirius is valid,
this does not mean that the same conclusion is valid for radiation from other stars.

Nevertheless we now consider what conclusions about the 
structure of the states $\Phi(t)$ can be made if we accept that
radiation from distant stars is classical.
As shown in textbooks, the operator of the vector potential has the form
\begin{equation}
A_{\mu}(x)=\sum_{\lambda}\int [exp(-\frac{i}{\hbar}px)e({\bf p},\lambda)_{\mu}a({\bf p},\lambda)+
exp(\frac{i}{\hbar}px)e({\bf p},\lambda)_{\mu}^*a({\bf p},\lambda)^*]\frac{d^3{\bf p}}{[(2\pi)^3|{\bf p}|]^{1/2}}
\end{equation}
where the four-vector $p$ has the component $(|{\bf p}|c,{\bf p})$, $\mu=0,1,2,3$ and $e({\bf p},\lambda)$ is the polarization vector of the photon with the momentum ${\bf p}$. Therefore the
antisymmetric tensor of the electromagnetic field can be written as
\begin{eqnarray}
&&F_{\mu\nu}(x)=\frac{\partial A_{\nu}(x)}{\partial x^{\mu}}-
\frac{\partial A_{\mu}(x)}{\partial x^{\nu}}=\nonumber\\
&&-i\sum_{\lambda}\int [exp(-\frac{i}{\hbar}px)
(p_{\nu}e({\bf p},\lambda)_{\mu}-p_{\mu}e({\bf p},\lambda)_{\nu})a({\bf p},\lambda)-\nonumber\\
&&exp(\frac{i}{\hbar}px)(p_{\nu}e({\bf p},\lambda)_{\mu}-p_{\mu}e({\bf p},\lambda)_{\nu})^*
a({\bf p},\lambda)^*]\frac{d^3{\bf p}}{[(2\pi)^3|{\bf p}|]^{1/2}}
\end{eqnarray}
Since in classical case the mean values of the operators $F_{\mu\nu}(x)$ should not be zero,
it follows from this expression and Eq. (\ref{commutatorA}) that classical states of the electromagnetic
field cannot be states with a fixed number of photons (this fact has been pointed out by the
referee of this paper); such states are complex superpositions of states with different numbers of photons.

However, this observation is not sufficient for making a conclusion on whether classical
states are coherent or noncoherent. In the literature it is often stated that the classical description
is valid for states with many photons while the quantum one is valid for states with
small numbers of photons. For pedagogical purposes it is often assumed that the electromagnetic field
is confined within a large finite volume and then the spectrum of momenta becomes discrete.
Then one can work with occupation numbers ${\bar n}({\bf p})$ characterizing mean values of photons
with the momentum ${\bf p}$. 

In these terms the question what states should be called classical or quantum is not
a matter of convention since in quantum theory there are rigorous criteria for that
purpose. In particular, as explained in textbooks on quantum theory, the exchange
interaction is a pure quantum phenomenon which does not have classical analogs.
That's why the Boltzmann statistics (which works when occupation numbers are
much less than unity and the exchange interaction is negligible) is classical while the
Fermi-Dirac and Bose-Einstein statistics (which work when occupation numbers
are of the order of unity or greater and the exchange interaction is important) are
quantum. 

So the criteria is not only whether the number of photons is large but also how 
this number is comparable to the number of possible states. In the case of the 
Boltzmann statistics the number of photons can be large but nevertheless it is
much less than the number of possible states, and the state $\Phi(t)$ 
is strongly incoherent. On the other hand, in lasers the number of photons is
much greater than the number of possible states. From the point of view of analogy with
the quantum mechanical oscillator problem the laser states are called coherent if the coefficients
$c_n$ in Eq. (\ref{coherent}) are such that 
\begin{equation}
a({\bf p},\lambda)\sum_{n=0}^{\infty}c_n[a({\bf p},\lambda)^*]^n\Phi_0=0
\end{equation}

Laser emission can be created only at very special conditions when energy
levels are inverted, the emission is amplified in the laser cavity etc. There are no reasons
to think that such conditions exist on stars. A part of the star
radiation consists of photons emitted from different atomic energy levels and this radiation
is fully spontaneous rather than induced. At the same
time, the main part of the radiation is understood such that it can
be approximately described in the blackbody model. 

As shown in textbooks, the occupation numbers for this radiation are given by
${\bar n}({\bf p})=1/[exp(E({\bf p})/kT)-1]$ where $E({\bf p})=|{\bf p}|c$, $k$ is the Boltzmann
constant and $T$ is the temperature of the blackbody radiation. If $E\gg kT$ then we get the
result for the Boltzmann statistics with zero chemical potential: ${\bar n}({\bf p})=exp(-kT/E)$.
The energy spectrum of the blackbody radiation has the maximum at $E/kT \approx 2.822$ \cite{LL5}.
 Hence in the
region of maximum ${\bar n}({\bf p})\approx 0.063$ and the result is close to that given by
the Boltzmann statistics.

When photons emitted by a star leave the area of the black body, their distribution differs from
the blackbody one. As argued, for example, in Ref. \cite{LL5}, this distribution can be described by the
Liouville theorem; in particular this implies that the photons leaving stars are moving
along classical trajectories. However, in any case, since the occupation numbers are small, it
is reasonable to think that $\Phi(t)$ is a superposition of states where photons are independent
of each other. 

In view of the above discussion, it is reasonable to think that the major part of radiation of stars
consists of free photons and therefore the effect of WPS for them can be considered for 
each photon independently.  

\section{Paradox: standard choice of position operator contradicts observations of stars}
\label{paradox}

As already noted, even if the function $f(\alpha)$ describes a broad 
angular distribution, a star will be visible only in the angular range of the order of
$R/L$. Hence one might think
that the absence of classical trajectories does not contradict observations. We now consider this
problem in greater details. For simplicity we first assume that the photon WF is spherically symmetric, i.e. $f({\bf r}/r)=const$.  

As already noted,  the WF of the photon coming to Earth
from a distant star is not negligible only within a thin sphere with the radius $ct$ and the width
of the order of $a$. On its way to Earth the sphere passes {\it all} stars, planets and other objects
the distance from which to the star is less than $L$ (in particular, even those objects which are from
the star in directions opposite to the direction to Earth). 
A problem arises how to explain the fact that the photon was detected on Earth and
escaped detection by those stars, planets etc. 

One might think that the event when the photon was detected on Earth is
purely probabilistic. The fact that the photon was not detected by the objects on its way
to Earth can be explained such that since the photon WF has a huge size (of the
order of light years or more) the probability of detection even by stars is extremely small
and so it was only a favorable accident that the photon was detected on Earth.

However, if the photon passed stars, planets
and other objects on its way to Earth then with approximately the same probability it can
pass Earth and can be detected on the opposite side of the Earth. In that case we could 
see stars even through the Earth.

Moreover, consider the following experiment. Suppose that we first look at a star and then place 
a small screen between the eye and the star. Then the experiment shows that the star will not
be visible. However, since the photon WF passed many big objects without
interacting with them then with approximately the same probability it can pass the screen.
In that case we could see the star through the screen. 

Those phenomena are not unusual 
in view of our understanding of neutrino physics.
It is known that neutrinos not only can pass the Earth practically without problems but even neutrinos created in the
center of the Sun can easily reach the Earth. The major neutrino detectors are
under the Earth surface and, for example, in the OPERA and ICARUS experiments
neutrinos created at CERN reached Gran Sasso (Italy) after traveling 730km under
the Earth surface. The explanation is that the probability of the neutrino interaction 
with particles comprising the Sun and the Earth is very small.

At low energies the electromagnetic interaction is much stronger than the weak one but,
as noted in Sec. \ref{WF}, the probability of interaction 
for photons having cosmic sizes contains the factor $|{\tilde f}/f|^2$. 
Therefore it is reasonable to expect that for such photons the 
probability of interaction with particles comprising an object is even much less
than in Earth experiments with neutrinos. 

In my discussions with physicists some of them proposed to avoid the above paradoxes by using an 
analogy with classical diffraction theory.  Here it is assumed that in optical phenomena
a wave falling on an object cannot penetrate inside the object. Then  
 the wave far from the object does not change, right after the object the wave has
a hole but when its length is much greater than the Rayleigh one the hole disappears and
the wave is practically the same as without diffraction. Those results 
are natural from the point of view
that classical waves consist of many almost pointlike particles. 

Let us now consider an experiment where a photon 
encounters a classical object and the transversal width of the photon coordinate WF
is much greater than the size of the object. By analogy with
diffraction theory one might represent
the photon WF as $\psi=\psi'+\psi''$ where the support of $\psi'$ is outside the
object and the support of $\psi''$ is inside the object. In contrast to diffraction theory, 
this decomposition is ambiguous because coordinates of the object have uncertainties. 
However, one might assume that the decomposition is valid with some accuracy.
Then one might expect that 
after interaction with the object $\psi'$ will not change and  $\psi''$ will be 
absorbed by the object. This statement can be formalized as follows.

Let the object be initially in the ground state $\Psi_g$. Then the initial WF of the system
photon+object is $\psi \Psi_g$. The S-matrix acts on this state as 
\begin{equation}
S(\psi \Psi_g)=S(\psi' \Psi_g)+S(\psi'' \Psi_g)=\psi' \Psi_g+(...)
\label{Greg}
\end{equation}
where (...) consists of states emerging after interaction. 
This expression describes the situation when the photon always interacts with the object but only a small 
part $\psi''$ of the initial WF interacts while the major part $\psi'$ remains intact.  
As a result of interaction, the photon
will be either absorbed by the object or will pass the object. In the latter case the photon
WF $\psi'$ will have a hole by analogy with the behavior of waves after diffraction.
Therefore in any case the photon cannot be detected in the geometrical shadow of the object.

Understanding whether or not Eq. (\ref{Greg}) is acceptable is crucial 
for drawing a conclusion on the above paradoxes. This expression can be justified if evolution is described by a Hamiltonian where interaction of the photon with the object is local. 
As already noted, in fundamental quantum theories elementary particles are described by 
states in the Fock space. 
If $\chi({\bf p})$ is the momentum WF then
the particle state in the Fock space is $\Phi_1=\int \chi({\bf p})a({\bf p})^*d^3{\bf p} \Phi_0$ where
for simplicity the polarization quantum numbers are suppressed.

If $\psi$ is the
coordinate WF defined by Eq. (\ref{Fourier}) then the decomposition $\psi=\psi'+\psi''$ corresponds
to the decomposition $\chi=\chi'+\chi''$ where 
\begin{equation}
\chi'({\bf p})=\int exp(-\frac{i}{\hbar}{\bf p}{\bf r})\psi'({\bf r})\frac{d^3{\bf r}}{(2\pi\hbar)^{3/2}},\quad
\chi''({\bf p})=\int exp(-\frac{i}{\hbar}{\bf p}{\bf r})\psi''({\bf r})\frac{d^3{\bf r}}{(2\pi\hbar)^{3/2}}
\label{chidecomp}
\end{equation}
Therefore the photon state in the Fock space can be represented as
\begin{equation}
\Phi_1=\int [\chi'({\bf p})+\chi''({\bf p})]a({\bf p})^*d^3{\bf p} \Phi_0
\label{Chi}
\end{equation}

In quantum mechanics particles exist during the whole time 
interval $t\in (-\infty,\infty)$ and evolution is defined by the interaction operator acting on 
the particle WF. This operator is local in 
coordinate space if in 
momentum space it acts on the particle WF $\chi$ as the convolution operator. However, if annihilation and creation of
particles are possible then evolution is described by operators acting not on particle WFs but on the operators
$a({\bf p})$ and $a({\bf p})^*$. In approximations when annihilation and creation is not important, evolution can be
reformulated in terms of $\chi$ only. For example, as noted in Sec. \ref{intro}, in the approximation $(v/c)^2$
the electron in the hydrogen atom can be described by the Dirac or Schr\"{o}dinger equation. 
 
However, Feynman diagrams contain only vertices with one photon. Therefore in any interaction the photon is first absorbed as a whole, in the intermediate state there is no photon, and in the case when the photon is reemitted this is a new photon. So in the case of interactions the evolution of the photon cannot be described
by an equation where the photon WF exists during the whole time interval $t\in (-\infty,\infty)$,
and the action of the evolution operator on photon states can be defined only in terms of $a({\bf p})$ and $a({\bf p})^*$. As follows from Eq. (\ref{commutatorA}), 
those operators are not local in coordinate space. So it is not possible that they act only on $\psi''$ and do not act on $\psi'$.

In general, if $\Psi$ is the WF of a system, and $\Psi=\Psi_1+\Psi_2$ is a decomposition of this function
then evolutions of $\Psi_1$ and $\Psi_2$ will be independent of each other if the states $\Psi_1$ and $\Psi_2$ have
at least one different conserved quantum number (e.g. angular momentum). However, in the decomposition 
$\psi \Psi_g=\psi' \Psi_g+\psi'' \Psi_g$ the
states are not characterized by a different conserved quantum number and therefore evolutions of the different parts of the decomposition will not be independent.

The crucial difference between diffraction theory and the given case follows. In diffraction theory
it is always known where different parts of the wave are. However, if a small part of the
photon WF is inside the object, this does not mean that a small part of the photon is inside
the object because the photon does not
have parts (roughly speaking, it is a point). Its WF describes only probabilities to find
the photon as a whole at different points. Hence  the fact that $\psi'' \neq 0$ does not mean that the part 
$\psi''$ of the photon is inside
the object but means only that the probability
to find the photon inside the object is not zero because this probability equals $||\psi''||^2$.
Since this quantity is very small then with the probability very close to unity the photon will
not interact with the object. 

This expectation is also in the spirit of QED. Since in any interaction the initial photon will be first
absorbed as a whole and there will be no photon in the intermediate state, the sizes of reemitted photons 
(if they are created) will be defined by the absorber, and after any interaction the WF of the object will 
not be $\Psi_g$. So in the case of interaction there is no part of the photon which does not participate in the interaction, and 
after any interaction WFs of final photons will not have large transverse sizes anymore. 

This is an illustration of the WF collapse: if the photon WF has a large size 
before interaction then, as a result of the WF collapse, after any interaction the WF cannot have a large size.
The WF collapse is a pure quantum phenomenon and there is no analog of WF collapse in
diffraction theory. 

A possible reason why Eq. (\ref{Greg}) might seem to be acceptable is that the
decomposition $\psi=\psi'+\psi''$ is implicitly (and erroneously) understood as breaking the photon
into two parts with the WFs $\psi'$ and $\psi''$. However, as noted in Sec. \ref{intro}, such a decomposition does not
mean that a particle is broken into two parts. Mathematically this is clear from the fact that 
the two-photon state
\begin{equation}
\Phi_{12}=const\int\int \chi'({\bf p}')\chi''({\bf p}'')a({\bf p}')^*a({\bf p}'')^*d^3{\bf p}'d^3{\bf p}'' \Phi_0
\label{Chi2}
\end{equation} 
fully differs from the state (\ref{Chi}).  

We conclude that the photon WF after interaction cannot be $\psi'$ and, instead of Eq. (\ref{Greg}), the result is
\begin{equation}
S(\psi \Psi_g)=c\psi \Psi_g+(...)
\label{Greg2}
\end{equation}
where $1-|c|^2$ is the probability of interaction. Since the probability is small, the quantity
$c$ is very close to unity and the photon will probably pass the objects without any
interaction. In rare cases when interaction happens, the WF of any final photon 
will not have a cosmic size anymore.  Such a photon can reach Earth only if its momentum 
considerably differs from the original one but this contradicts Theory A. So the assumption that the above 
paradoxes can be explained by analogy with diffraction theory is not justified. 

If $f({\bf r}/r)\neq const$ then, as follows from Eqs. (\ref{FG}) and (\ref{relpsif}),
the radial part of the WF is the same as in the spherically symmetric case and, 
as follows from the above discussion, the coordinate WF of the initial photon still has a cosmic size.
Therefore on its way to Earth the photon WF will also encounter stars, planets and
other objects (even if they are far from the line connecting the star and Earth) 
and the same inconsistencies arise.

{\it In summary, since according to standard theory photons emitted by stars have coordinate wave functions with cosmic sizes,
the above arguments indicate that the theory contradicts observational data.}

\section{The reason of the paradox and its resolution}
\label{reason}

As noted in Sec. \ref{intro}, {\it the results} of fundamental quantum theories are formulated only in momentum space without mentioning space-time at all. So for those theories the 
position operator is not needed. It is needed only in semiclassical approximation, and, as noted
in Sec. \ref{intro}, the choice $i\hbar \partial/\partial {\bf p}$ is often substantiated 
from the requirement of compatibility with this approximation. 
Let A be a physical quantity, ${\bar A}$ be its mean value in some state and $\Delta A$ be
its uncertainty in this state. Then $A$ is semiclassical if $\Delta A\ll 
|{\bar A}|$. In particular, $A$ cannot be semiclassical if ${\bar A}=0$.

Consider first the one-dimensional case. As argued in textbooks (see e.g. Ref. \cite{LLIII}),
if the mean value of the $x$ component of the momentum $p_x$ is rather large, the operator 
$i\hbar \partial/\partial p_x$ can be semiclassical  but it is not semiclassical in situations 
when $p_x$ is small. 
This is clear even from the fact that if $p_x$ is small then $exp(ip_xx/\hbar)$ is not a rapidly oscillating function of $x$.

Consider now the three-dimensional case.
If all the components $p_j$ ($j=1,2,3$) are rather large then all the operators 
$i\hbar \partial/\partial p_j$ can be semiclassical. A semiclassical wave function $\chi({\bf p})$ in
momentum space describes a narrow distribution around the mean value ${\bf p}_0$. Suppose  that coordinate 
axes are chosen such ${\bf p}_0$ is directed along the $z$ axis. Then the mean values of the $x$ and $y$ components of the momentum operator equal zero and
the operators $i\hbar \partial/\partial p_j$ cannot be semiclassical for $j=1,2$, i.e. in directions perpendicular to the particle momentum. 
 Hence standard definition of the transverse components of the position operator is not 
semiclassical.

The effect of WPS in transverse directions discussed in Sec. \ref{WF} is a consequence of
standard definition of the position operator in these directions. As shown in Sec. \ref{paradox},
this results in fundamental paradoxes in observations of stars. Therefore the inconsistent 
definition of the position operator in these directions is not a pure academic problem.

In Ref. \cite{lev1} a new definition of the position operator has been proposed. In contrast to standard definition, the new operator does not depend on the choice of coordinate axes and is expected
to be physical only if the {\it magnitude} of the momentum is rather large. As a consequence, 
WPS in directions perpendicular to the particle
momentum is absent regardless of whether the particle is nonrelativistic or relativistic. 
Moreover, for an ultrarelativistic particle the effect of WPS is absent at all. 

In this approach uncertainties of each component of the photon momentum and each component of the photon coordinate do not 
change with time. If those quantities can be treated as small then the photon can be (approximately) treated as  a pointlike 
particle moving along classical trajectory. 
So uncertainties of photon coordinates never have a cosmic size and there
can be no paradoxes discussed in Ref. \cite{lev1} and Sec. \ref{paradox}. 

As noted in Secs. \ref{intro} and \ref{paradox}, a very important point in understanding the paradoxes is that the wave 
function describes only amplitudes of probabilities. This is important in discussing other quantum phenomena.
As an example, consider the double-slit experiment which is usually treated as a strong
confirmation of standard quantum theory. A standard explanation is that parts of the wave
function of an elementary particle projected on the slits pass the screen and interfere, and
the remaining part is absorbed by the screen. However, in view of the above consideration, it is not consistent to
treat the elementary particle by analogy with the classical wave different parts of which interact with the screen differently.
The problem of understanding the experiment is very difficult because here the WF of the elementary particle
does not have an anomalously large size and the particle strongly interacts with the screen.

\begin{center} {\bf Acknowledgments} \end{center}
I am very grateful to Gregory Keaton and Anatoly Kamchatnov for numerous important discussions and constructive criticism.  Remarks of the (anonymous) referee were very important for improving the
paper. In particular, those remarks prompted me to add Section 3.
 I am also grateful to Volodya Netchitailo for discussions.

\end{document}